\documentclass[pra,a4,
              twocolumn,
              showpacs,amsmath,amssymb]{revtex4-1}

\usepackage{graphicx}
\usepackage{graphicx,subfigure,epsfig}
\usepackage{color}
\usepackage{soul}
\usepackage{float}
\usepackage{amsmath}
\usepackage{amssymb}
\usepackage{ifthen}
\usepackage{alltt}
\usepackage{fancyhdr}
\usepackage{verbatim}
\usepackage{moreverb}
\usepackage{colortbl}
\usepackage{multirow}
\usepackage{bigstrut}
\usepackage{wrapfig}
\usepackage{enumerate}
\usepackage{rotating}
\usepackage{epstopdf}

\begin{document}

\title{Double resonant wideband Purcell effect in a polaritonic wire medium}
\author{M.~S.~Mirmoosa$^1$, S.~Yu.~Kosulnikov$^{1,2}$ and C.~R.~Simovski$^{1,2}$}
\affiliation{$^1$Department of Radio Science and Engineering, School of Electrical Engineering, Aalto University, P.O.~Box 13000, FI-00076 AALTO, Finland\\
$^2$Department of Nanophotonics and Metamaterials, ITMO University, 197043, St.~Petersburg, Russia}
\date{\today }

%%%%%%%%%%%%%%%%%%%%%%%%%%%%%%%%%%%%%%%%%%%%%%%%%%%%%%%%%%%%%%%%%%%%%%%%%%%%%%%%%%%%%%%%%%%%%%%%%%%%%%%%%%%%

\begin{abstract}
In this paper, we theoretically show that a broadband resonant enhancement of emission may occur for infrared sources located in a polaritonic wire medium. The reason of this enhancement is overlapping of two topological transitions of the wave dispersion in the medium. The first topological transition has been recently revealed as an effect inherent to polaritonic wire media. The second one was theoretically uncovered in another material. In this work we reveal it for wire media and prove the possibility to combine both these transitions with the purpose to obtain the broadband resonant Purcell factor. We compare the results obtained for two orientations of a subwavelength electric dipole embedded into wire medium -- that along the optical axis and that perpendicular to it -- and report on the resonant isotropic radiation enhancement. Also, we reveal the enhancement of radiation to free space from a finite sample of the wire medium.
\end{abstract}

\pacs{41.20.Jb, 81.05.Xj, 78.67.Uh, 74.25.Gz}

\maketitle

%%%%%%%%%%%%%%%%%%%%%%%%%%%%%%%%%%%%%%%%%%%%%%%%%%%%%%%%%%%%%%%%%%%%%%%%%%%%%%%%%%

\section{introduction}
Throughout the last decade, enormous efforts have been put into research and development of artificial electromagnetic media with unusual properties. One of these media is the so-called wire medium (WM), defined as an optically dense array of parallel wires embedded in a host dielectric matrix, as depicted in Fig.~\ref{fig:medium}. Its unique feature is applicability from radio frequencies to those of the visible light (see, e.g., Ref.~\cite{simovski2}). Subwavelength imaging \cite{belov1, casse}, improvement of thermophotovoltaic systems \cite{simovski1, mirmoosa1}, enhancement of directionality of microwave antennas \cite{burghignoli1} are some examples of such applications. One of prominent applications is optical sensing and location of fluorescent nano-objects at optical frequencies. Generally speaking, this possibility is provided because the power radiated by the fluorescent emitter located in such a medium may increase several times or even by an order of magnitude (see, e.g., Ref.~\cite{poddubny3}).

More in detail, since wire medium is optically dense, it can be homogenized, and based on the effective medium model, it is a uniaxially anisotropic material whose permittivity tensor has principal components which (at least at some frequencies) have different signs \cite{simovski2}. Therefore, the sign of the permittivity tensor trace is indefinite, and wire medium can be classified as an indefinite material according to Ref.~\cite{smith1}. This feature affects remarkably the isofrequency surface (dispersion surface in the reciprocal lattice called $\mathbf{k}$-space) of the extraordinary modes propagating in the medium, whereas the isofrequency for the ordinary mode keeps similar to that of an isotropic medium. The indefinite regime implies an open surface close to a hyperboloid \cite{simovski2}. Hence, at corresponding frequencies, wire medium refers to the class of so-called hyperbolic metamaterial (see, e.g., Ref.~\cite{poddubny1}). The hyperbolic topology implies very high spatial frequencies of propagating extraordinary modes and, respectively, very high density of electromagnetic states. This effect results in the enhancement of the radiated power of a dipole source submerged into the medium (see, e.g., Refs.~\cite{poddubny2, kidwai1}). Increase of the radiated power implies enhancement of spontaneous emission and decay rate. The enhancement of the decay rate due to an environment of an emitter is called Purcell's effect. Historically, this effect was referred as the enhanced decay rate of the excited state of a quantum emitter due to insertion of this emitter into an open cavity \cite{purcell1, novotny1}. However, recently the concept of Purcell's effect was generalized to any resonant scatterer located in the vicinity of the emitter \cite{sauvan1, pelton1, tam1, anger1}, and finally -- to the case of an arbitrary environment different from free space (see, e.g., Refs.~\cite{jacob1,poddubny3,Krasnok}). The increase of the radiated power of a monochromatic point dipole due to its environment is called the Purcell factor. Consequently, WM may offer a high Purcell factor in the hyperbolic regime. 

High Purcell factor is inherent to any hyperbolic material. A periodic stack of strongly submicron bilayers, where one layer is a plasmonic material and another is a material with positive permittivity, is a popular implementation of a hyperbolic medium operating in the optical frequency range (see, e.g., Refs.~\cite{poddubny1,krishnamoorthy1}). Therefore, this medium also allows a high Purcell factor \cite{jacob1}. Moreover, the Purcell factor for such medium can be resonant, i.e. attains a maximum at a certain frequency. This resonance is claimed for the case when a transition happens in the topology of the isofrequency surface. Namely, the resonant maximum of the Purcell factor occurs at a frequency where the isofrequency of the stacked hyperbolic medium transits from the closed-surface (elliptical) regime to open-surface (hyperbolic) regime or vice versa \cite{krishnamoorthy2}. This transition was called topological phase transition \cite{krishnamoorthy2, yang}.

In our recent work \cite{new}, we studied the resonat behavior of the Purcell factor for a wire medium formed by nanowires prepared from a so-called polaritonic material and operating at infrared frequencies. The resonance corresponds also to the topological transition. However, the topological transition in this medium, unlike that of the stacked material, offers an unbounded spatial spectrum of radiation to internal sources. At the topological transition revealed in Ref.~\cite{new}, the isofrequency surface is composed of both the hyperbolic and elliptic dispersion branches connected with one another as depicted in Fig.~\ref{fig:tran1}. Since the spatial spectrum is unbounded over the parallel wave vector $\beta$ (vertical axis in the Fig.~\ref{fig:tran1}), a parallel dipole generates more radiation than the perpendicular one \cite{new}. This result makes our work Ref.~\cite{new} actual in view of the aforementioned application.

In the present paper, we report that the same polaritonic WM as was studied in Ref.~\cite{new} experiences, in fact, several topological transitions. Two of them are important since correspond to the resonant enhancement of the dipole radiation. Restricting the frequency band of our interest, we have not noticed the second useful transition in Ref.~\cite{new}. In fact, the two transitions occur at neighboring frequencies. On some feasible conditions, two corresponding resonances of the Purcell factor may overlap. This overlapping has an important implication for the radiation of dipole sources in a polaritonic WM. Both values and frequency dependencies of the Purcell factor turn out to be very different from those predicted in the literature for a stacked hyperbolic material. This difference is favorable for a WM and makes it, to our opinion, more attractive for infrared nanosensing than the stacked hyperbolic materials. This is the main message of the present work.

The paper is organized as follows: in Sec.~\ref{sec:theory} we discuss these two topological transitions, in Section~\ref{sec:results} we summarize the results, and finally in Sec.~\ref{sec:conclusions}, we conclude the paper.

%%%%%%%%%%%%%%%%%%%%%%%%%%%%%%%%%%%%%%%%%%%%%%%%%%%%%%%%%%%%%%%%%%%%%%%%%%%%%%%%%%

\section{Theory}
\label{sec:theory}
\begin{figure}[t!]\centering
\subfigure[]{
\includegraphics[width=8cm]{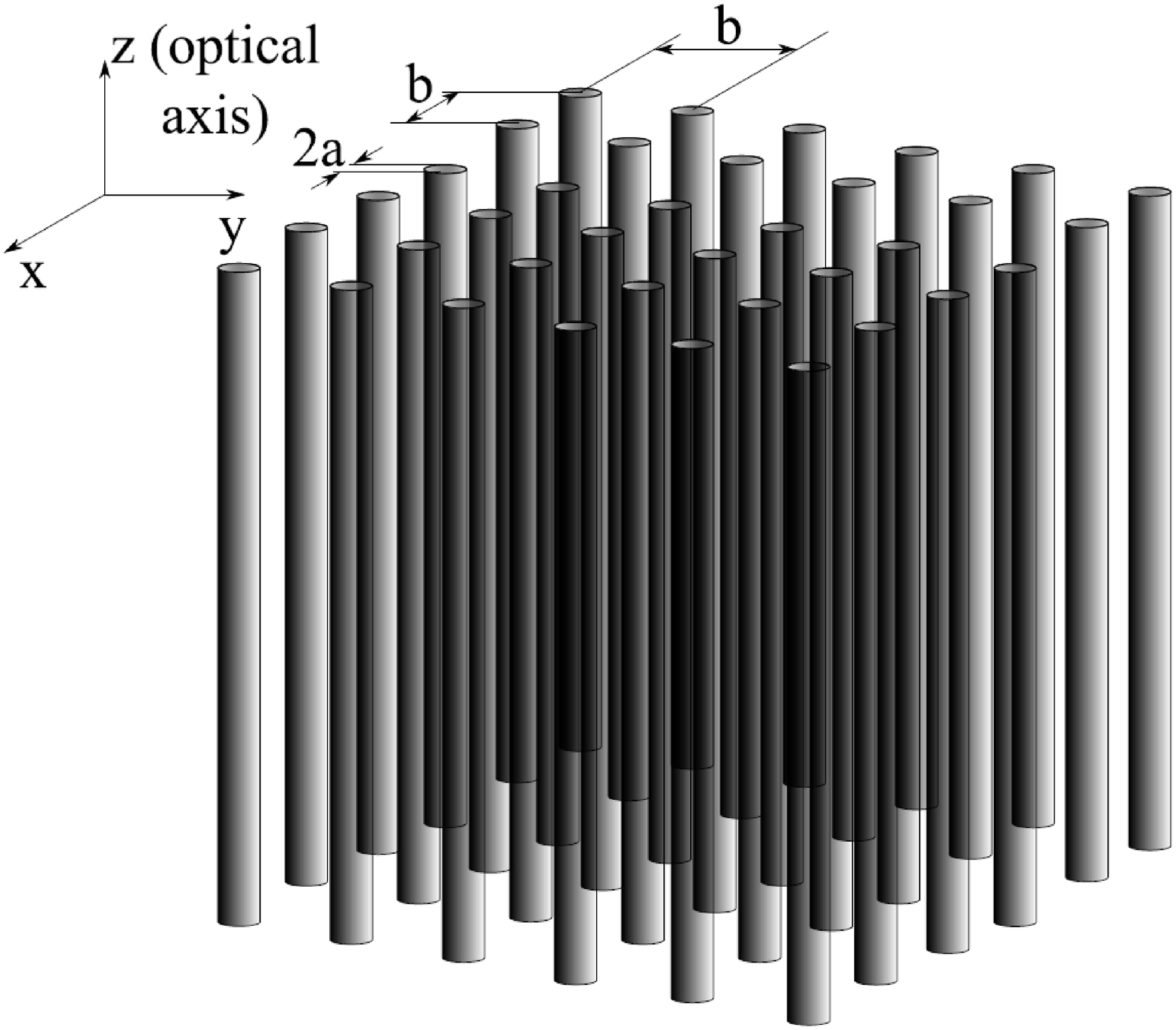}
\label{fig:medium}}
\subfigure[]{
\hspace*{0cm}\includegraphics[width=2.8cm]{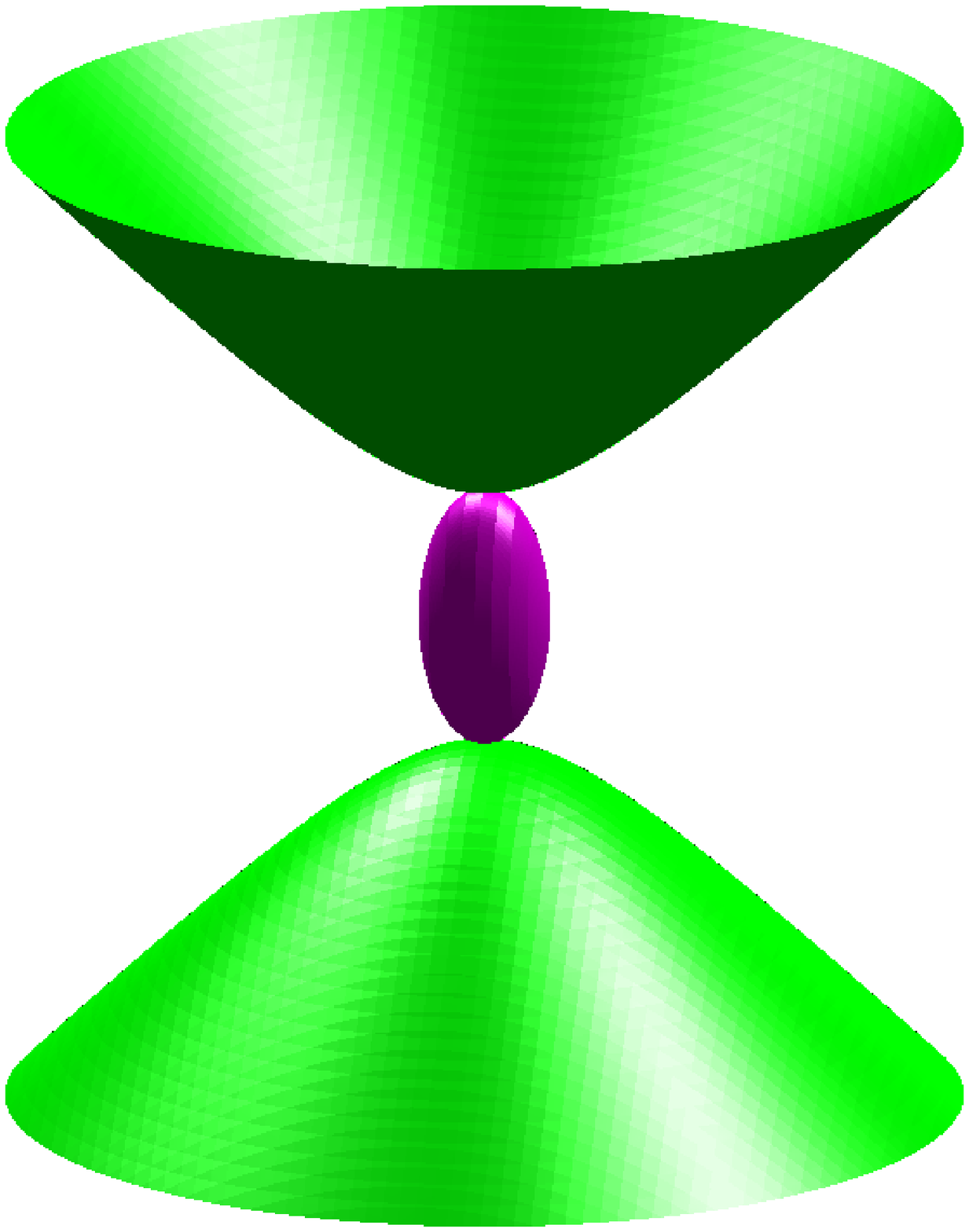}
\label{fig:tran1}}
\subfigure[]{
\hspace*{1cm}\includegraphics[width=4.3cm]{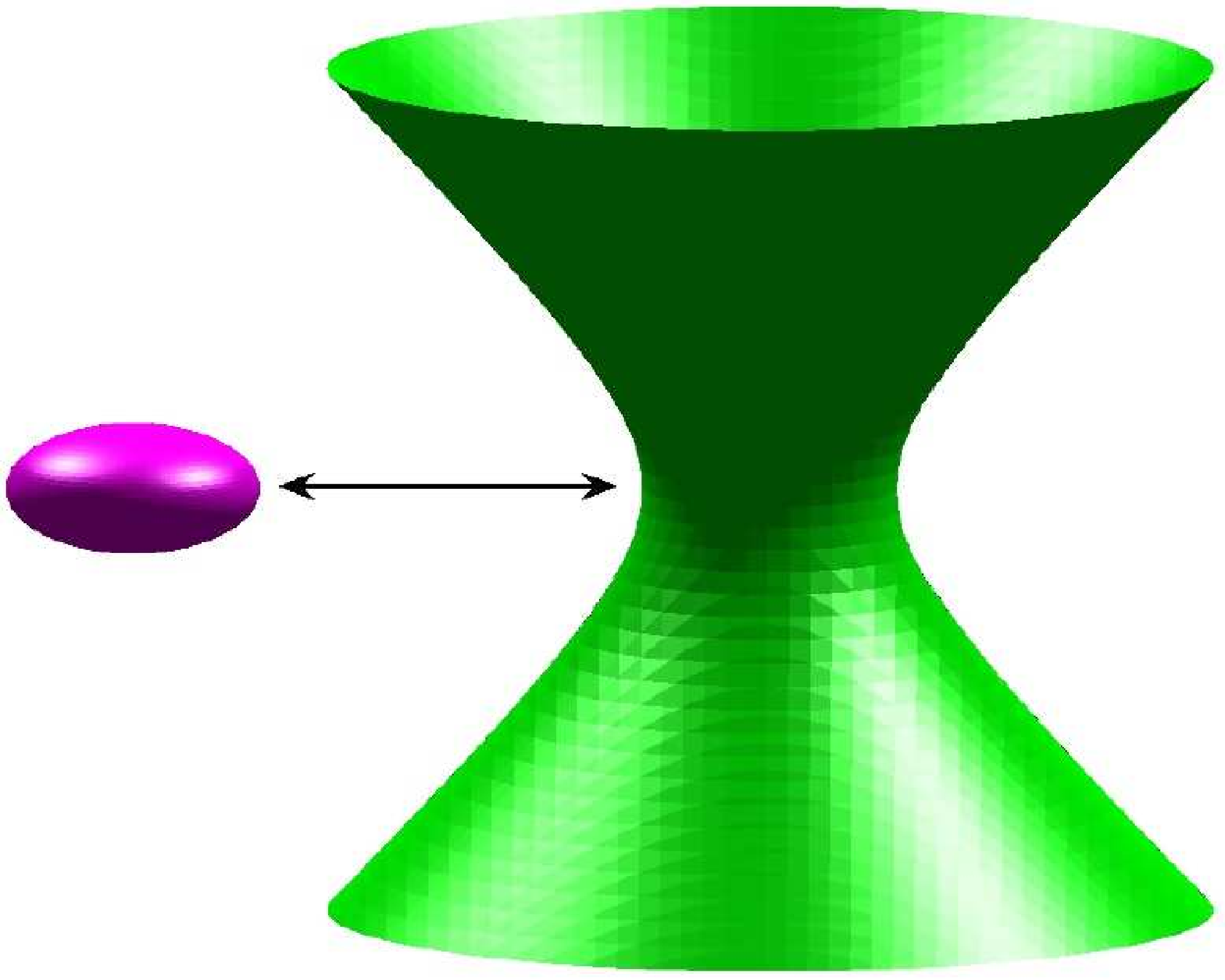}
\label{fig:tran2}}
\caption{(a)--A medium of parallel wires which form a square lattice. (b)--Isofrequency of the first useful topological transition comprises both closed-surface (elliptic) and open-surface
(two-branch hyperbolic) surfaces. (c)--The second useful topological transition occurs as a sharp (in the lossless case) substitution of an ellipsoid by a solid hyperboloid.}
\end{figure}

Figure~\ref{fig:medium} schematically illustrates a WM as a regular optically dense array of parallel cylindrical wires embedded in a host dielectric material. In our calculations, we assume for simplicity that the host material is free space.
As the figure shows, the optical axis of the medium has unit vector $\mathbf{z}_0$ which is also the wire axis. In this case, homogeneous effective permittivity tensor of the medium is given by
\begin{equation}
\overline{\overline{\epsilon}}=\left(\begin{array}{ccc}
\varepsilon_{xx} & 0 & 0 \\
0 &\varepsilon_{yy} & 0 \\
0& 0 & \varepsilon_{zz} \end{array} \right)=\left(\begin{array}{ccc}
\varepsilon_{\perp} & 0 & 0 \\
0 &\varepsilon_{\perp} & 0 \\
0& 0 & \varepsilon_{\parallel} \end{array} \right),
\end{equation}
in which \cite{silveirinha1}
\begin{equation}
\begin{split}
&\varepsilon_\perp=1+\displaystyle\frac{2}{\displaystyle\frac{\varepsilon_{\rm{r}}+1}{f_{\rm{v}}(\varepsilon_{\rm{r}}-1)}-1},\\
&\varepsilon_\parallel=1+\displaystyle\frac{1}{\displaystyle\frac{1}{f_{\rm{v}}(\varepsilon_{\rm{r}}-1)}-\displaystyle\frac{k_0^2-\beta^2}{k_{\rm{p}}^2}}.
\end{split}
\label{eq:effectivep}
\end{equation}
The subscripts $\perp$ and $\parallel$ mean components, perpendicular and parallel to the optical axis, respectively. In Eq.~\ref{eq:effectivep}, $f_{\rm{v}}$ is the volume fraction and $\varepsilon_{\rm{r}}$ represents the relative dielectric constant of the wires. Furthermore, $k_0$ and $k_{\rm{p}}$ are the free-space and plasma wave numbers, respectively, and $\beta$ is the parallel component of the wave vector. The effective-medium model \cite{silveirinha1} was developed for wires performed of a material with negative dielectric response. Its applicability is determined by two main conditions -- that of sufficiently low spatial frequencies $qb\leq\pi$ ($q$ is the perpendicular component of the wave vector) and that of long wavelengths $k_0b\ll2\pi$. If the radius of the wires is much smaller than the wavelength ($k_0a\ll1$) and the fraction of wires is rather small (practically $f_{\rm{v}}<0.3$), the effective-medium model \cite{silveirinha1}, in accordance to several recent studies, is quite accurate.

Distinguished by directions of the electric and magnetic fields, two different modes exist in the WM. The first one is transverse electric (TE) or ordinary mode, whose electric field is normal to the optical axis. The dispersion equation for this mode is standard:
\begin{equation}
q^2+\beta^2=k_0^2\varepsilon_\perp.
\end{equation}
The isofrequency is a simple sphere with the radius $\sqrt{\varepsilon_\perp}$.
If the fraction volume is small and the relative dielectric constant of the wires is not close to $-1$, the perpendicular component of the effective permittivity is close to unity. It implies that the TE modes do not interact with the wires and the Purcell effect for a source creating TE-polarized radiation is absent.

The second eigenmode is transverse magnetic (TM) or extraordinary mode whose magnetic field is normal to the optical axis. Its dispersion equation can be written as follows (see, e.g., Ref.~\cite{marcuvitz1}):
\begin{equation}
\displaystyle\frac{q^2}{\varepsilon_\parallel}+\displaystyle\frac{\beta^2}{\varepsilon_\perp}=k_0^2.
\label{eq:dis}
\end{equation}
As mentioned in the Introduction and Eq.~\ref{eq:dis} determines, depending on the sign of $\varepsilon_\parallel$ and $\varepsilon_\perp$ (more exactly, the real parts of these components), an open surface or a closed surface arises in the reciprocal space. Definitely, $\Re{(\varepsilon_\parallel)}<0\,\&\,\Re{(\varepsilon_\perp)}>0$, or $\Re{(\varepsilon_\parallel)}>0$\,\&\,$\Re{(\varepsilon_\perp)}<0$ gives an open surface, and $\Re{(\varepsilon_\parallel)}>0$\,\&\,$\Re{(\varepsilon_\perp)}>0$ results in a closed isofrequency surface. These surfaces can be different from, respectively, hyperboloid and ellipsoid due to the spatial dispersion (dependence of the effective permittivity on $\beta$).

In the following subsections, we discuss two different types of transitions from one dispersion surface to another one. At the first type of transition, the two-branch hyperboloid transits versus growing frequency to an ellipsoid. This transition in polaritonic WM comprises several stages which may be in their turn called separate topological transitions. At low frequencies the isofrequency surface of our WM is a two-branch hyperboloid. Starting from some frequency there appears an small ellipsoid which enlarges versus frequency and coexists with the hyperboloid. This is the first stage of the topological transition studied in Ref.~\cite{new}. The second stage corresponds to the certain frequency at which both hyperbolic and elliptic surfaces connect one another and form a solid isofrequency. At the third stage the cut-off of the hyperboloid in the reciprocal space increases and it disconnects from the ellipsoid. This growth of the cut-off versus frequency is very fast, and at some frequency the hyperboloid disappears from the domain of physically sound values of the wave vector. This dynamics of the isofrequency surface can be treated as a multi-stage transition from the hyperbolic to the elliptic dispersion of propagating waves. However, in Ref.~\cite{new}, we mainly concentrated on the stage when the hyperbolic and elliptic surfaces are connected, since it grants the unbounded spatial spectrum of eignemodes. This regime grants the strongest enhancement to the dipole radiation and we observed the maximum of simulated Purcell factor namely at this frequency. 

At the second topological transition, illustrated by Fig.~\ref{fig:tran2}, an ellipsoid is substituted by a solid hyperboloid. This hyperboloid describes the dispersion in uniaxial media whose perpendicular permittivity is negative whereas the parallel one is positive. Below we show that this is so for polaritonic WM at sufficiently high frequencies. Therefore, both types of topological transition hold in the same material - WM of polaritonic nanowires. 

Notice that in Ref.~\cite{poddubny3}, the resonance of the Purcell factor of a WM performed of $\varepsilon_{\rm{r}}$-negative nanowires was revealed, though not explained. We have calculated isofrequencies for WM whose Purcell factor is depicted in Fig.~4 of Ref.~\cite{poddubny3} and found that the resonance of the Purcell factor for a parallel dipole (which occurs at $\varepsilon_{\rm{r}}=-90$, $k_0b/\pi=a/b=0.05$) corresponds namely to our topological transition of the first type.

\subsection{Topological transition via unbounded spatial spectrum}

Though this transition has been studied in Ref.~\cite{new}, in this subsection, we briefly review this work for the convenience of the reader. 

Equation~\ref{eq:effectivep}, where we adopt for simplicity an approximation $\varepsilon_\perp\approx1$, together with Eq.~\ref{eq:dis} delivers two solutions for TM-waves:
\begin{equation}
\beta^2=k_0^2-\displaystyle\frac{q^2+k_{\rm{p}}^2-k_{\rm{c}}^2\pm\displaystyle\sqrt{(q^2+k_{\rm{p}}^2-k_{\rm{c}}^2)^2+4k_{\rm{c}}^2q^2}}{2}.
\label{eq:beta}
\end{equation}
Here it is denoted:
\begin{equation}
k_{\rm{c}}^2=\displaystyle\frac{k_{\rm{p}}^2}{f_{\rm{v}}(1-\varepsilon_{\rm{r}})}.
\end{equation}
 
The first solution corresponds to open-surface dispersion ($\varepsilon_\parallel<0$), and the second solution -- to closed-surface dispersion ($\varepsilon_\parallel>0$). The separation of these two surfaces in the reciprocal space is determined by $k_{\rm c}$.  
If $|\varepsilon_{\rm{r}}|$ is very large, $k_{\rm{c}}$ is much smaller than the plasma wave number $k_{\rm{p}}$. In this scenario, $\varepsilon_\parallel$ is negative at the corresponding frequency and the isofrquency surface is open suffering a cut-off over the parallel component of the wave-vector axis: $\vert\beta\vert\geq k_0$. The closed surface in the reciprocal space is absent since the second solution in Eq.~\ref{eq:beta} corresponds to $\beta^2<0$. However, if $\vert\varepsilon_{\rm{r}}\vert$ is a modest value, the parameter $k_{\rm{c}}$ becomes comparable to $k_{\rm{p}}$. Now, the second solution like the first one also corresponds to $\beta^2>0$. Therefore, the closed surface also appears and both surfaces coexist as parts of the complex isofrequency surface. Note that the closed surface covers low values of the wave vector: $\vert\beta\vert<k_0$. Finally, at the frequency where $k_{\rm{c}}=k_{\rm{p}}$, the first ($\vert\beta\vert\geq k_0$) and the second ($\vert\beta\vert<k_0$) solutions touch each other at $\beta=\pm k_0$ and the surfaces are connected. This connection gives topological phase transition of the first type which is the most important stage of this multi-stage process because the spatial spectrum becomes unbounded. Not in any WM the regime with transition of $k_{\rm{c}}$ can approach to $k_{\rm{p}}$ is achievable, however, such WM exists.

At the points $\beta=\pm k_0$ where the open and closed surfaces are connected, the parallel component of the effective permittivity is zero $\varepsilon_\parallel=0$. As mentioned, $k_{\rm{c}}=k_{\rm{p}}$, therefore, after some easy algebra, we obtain the condition of this desired regime in the form
\begin{equation}
\varepsilon_{\rm{r}}=1-\displaystyle\frac{1}{f_{\rm{v}}}.
\label{eq:condition}
\end{equation}
This condition is sufficient for the topological transition of the first type in the framework of the effective-medium model \cite{silveirinha1}. 

%This model implies $f_{\rm{v}}\ll 1$, however, $f_{\rm{v}}$ cannot be negligibly small because the nanowires must be substantial enough. Therefore, the required permittivity of nanowires %is modestly negative. 
%We have searched the first topological transition for WM of plasmonic nanowires (Au, Ag, Al) operating in the visible frequency range. However, this regime either required design %parameters of nanowires not yet reported in the literature (too challenging for fabrication) or is not compatible with the effective-medium model. Therefore, in this work we still concentrate %on polaritonic nanowires operating in the infrared range. 

Suppose that the frequency dependence of the dielectric response of the wires can be modeled as
\begin{equation}
\varepsilon_{\rm{r}}({\rm{\omega}})=\varepsilon^{(0)}-\displaystyle\frac{K_1}{k_0-K_2-jK_3},
\label{eq:model}
\end{equation}
For low-loss materials, we may approximately equate right-hand sides of  Eq.~\ref{eq:condition} and Eq.~\ref{eq:model}. Optical losses in the material of nanowires, if sufficiently small, are irrelevant for this topological transition, and in this subsection we neglect them. Of course, in our work \cite{new} we took these losses into account and considered their influence.

\begin{figure}[htb!]\centering
\includegraphics[width=8.5cm]{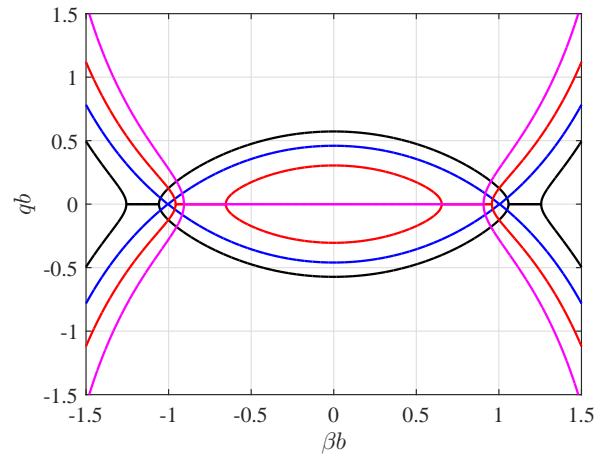}
\caption{Isofrequency contours (cross section of the dispersion surface) for TM mode. Calculations are done based on the following values: $a/b=0.05$, $K_1b=50$, $K_2b=0.6073$, $K_3b=0$ and $\varepsilon^{(0)}=1$.
Magenta color corresponds to normalized frequency $k_0b=0.90$, red color -- to $k_0b=0.95$, blue -- to $k_0b=1$, and black -- to $k_0b=1.05$.}
\label{fig:hybrid}
\end{figure}

In Fig.~\ref{fig:hybrid}, the normalized perpendicular wave vector is plotted versus the normalized parallel wave vector for four normalized frequencies $k_0b=0.9$, $k_0b=0.95$, $k_0b=1$, and $k_0b=1.05$.
In these calculations, we put $a/b=0.05$, $\varepsilon^{(0)}=1$, $K_1b=50$, $K_2b=0.6073$, and $K_3=0$. 
At~$k_0b=0.9$, the isofrequency surface (magenta color) is open and close to a hyperboloid. This frequency corresponds to  $\varepsilon_{\rm{r}}=-169.82$ and $k_{\rm{c}}/k_{\rm{p}}=0.8633$. At higher frequencies $|\varepsilon_{\rm{r}}|$ decreases and $k_{\rm{c}}$ approaches $k_{\rm{p}}$. At $k_0b=0.95$ (the first stage of the first-type topological transition) the open and closed branches coexist but are separated. Their touching occurs at $k_0b=1$ when $\varepsilon_{\rm{r}}=1-1/f_{\rm{v}}=-126.324$
and $k_{\rm{c}}=k_{\rm{p}}$. At this frequency the spatial spectrum is unbounded. This is the second and most important stage of the first-type topological transition. At $k_0b>1$, $\vert\varepsilon_{\rm{r}}\vert$ further decreases and the two surfaces separate again (the third stage). At a higher frequency the hyperbolic branch disappears and only the elliptic one remains (the fourth stage).

\subsection{Topological transition from ellipsoid to solid hyperboloid}

If we consider an extended frequency range, we may notice that the another topological transition occurs in the same WM.
Really, based on Eq.~\ref{eq:effectivep}, we may write
\begin{equation}
\varepsilon_{\perp}=-\displaystyle\frac{1}{Z}\displaystyle\frac{\varepsilon_{\rm{r}}(\omega)-Z}{\varepsilon_{\rm{r}}(\omega)-P},
\label{eq:tper}
\end{equation}
where the pole $P$ and zero $Z$ are expressed as
\begin{equation}
P=-\displaystyle\frac{1+f_{\rm{v}}}{1-f_{\rm{v}}},\,\,\,\,\,\,\,\,Z=-\displaystyle\frac{1-f_{\rm{v}}}{1+f_{\rm{v}}}.
\label{eq:pz}
\end{equation}
Both the zero and the pole of $\varepsilon_\perp$ exist at higher frequencies than that of the first topological transition.
As the above equation indicates, $P<-1$ and $-1<Z<0$. If the fraction of nanowires is very small, the pole and zero are close to each other on the frequency axis. Definitely, at the pole ($\varepsilon_{\rm{r}}=P$) and at the zero ($\varepsilon_{\rm{r}}=Z$) the sign of the perpendicular component $\varepsilon_\perp$ changes. Recall, that the first topological transition occurs at a frequency where ${\rm Re}(\varepsilon_\parallel)$ may vanish (vanishes when $|\beta|=k_0$). The zero of ${\rm Re}(\varepsilon_\perp)$ also results in a topological transition. This second transition exploits the frequency dispersion of $\varepsilon_{\rm{r}}$. The spatial dispersion is here insignificant -- this zero occurs at the corresponding frequency for any $\beta$.

Definitely, the hyperbolic isofrequency extends to infinity and therefore implies the broader spatial spectrum than the elliptic one. Based on this one could expect the growing Purcell factor at frequencies above this topological transition. However, in fact, a local maximum of Purcell fcator occurs exactly at this frequency. This maximum was explained in Ref.~\cite{chebykin} for the case of a stacked hyperbolic medium using the previously obtained Purcell factor of a uniaxial dielectric. Really, the zero of the perpendicular effective permittivity, which occurs at the topological transition of the second type (whereas the parallel permittivity keeps finite), implies the ultimate anisotropy of the medium. Consequently, the radiation of any finite source (whatever small) should be super-directional. The super-directionality of the dipole radiation implies the singularity for the density of photonic states, and the Purcell factor becomes maximally high. If both parallel and perpendicular components of the effective-medium permittivity tensor are positive the formula for the parallel-dipole Purcell factor describing this effect takes form \cite{chebykin}:
\begin{equation}
P_{\rm{F}}={\varepsilon_\parallel\over 4\sqrt{\varepsilon_\perp}}+3{\sqrt{\varepsilon_\perp}\over 4}.
\label{eq:FP}
\end{equation}
This estimation gives an the infinite value of the Purcell factor in the lossless case, because does not take into account the limit of the effective-medium model for spatial frequencies. 
Definitely, this formula is not applicable for lossy hyperbolic media, it only gives an insight of the effect of ultimate anisotropy $\varepsilon_\parallel\gg \varepsilon_\perp$.  

\begin{figure}[t!]\centering
\includegraphics[width=8.5cm]{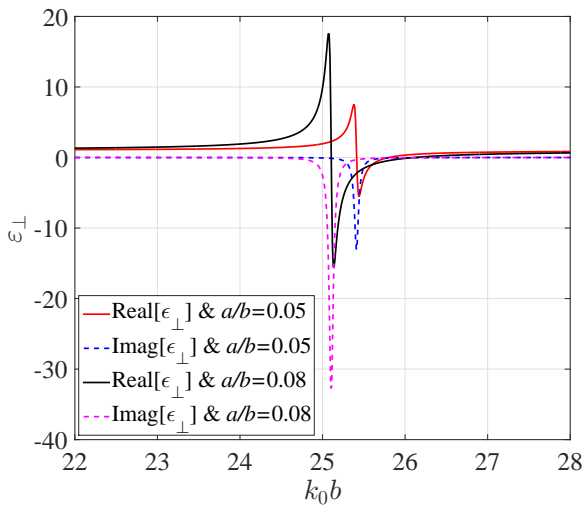}
\caption{Perpendicular component of the effective permittivity (real and imaginary parts -- solid and dashed, respectively) versus normalized frequency.
Red and black curves correspond to $a/b=0.05$ and $a/b=0.08$, respectively. Calculations are done for the case $\varepsilon^{(0)}=1$, $K_1b=50$, $K_2b=0.6073$ and $K_3b=0.03$.}
\label{fig:e_perp}
\end{figure}

Fig.~\ref{fig:e_perp} shows $\varepsilon_\perp$ versus frequency, where Eq.~\ref{eq:model} describes the frequency dispersion of the material of nanowires. In the present calculations, the losses in this material are taken into account: parameters of the material are chosen as follows: $\varepsilon^{(0)}=1$, $K_1b=50$, $K_2b=0.6073$ (as before), and $K_3b=0.03$. Optical losses transform the pole and the zero of the perpendicular permittivity into a Lorentz's type resonance. The pole is substituted by the maximum of $|\varepsilon_\perp|$ and the zero refers in the lossy case to the real part. An increase of the volume fraction of nanowires causes the red shift of the resonance frequency $\varepsilon_\perp$ and enhances the magnitude of the resonance.
Notice that in the whole frequency range of Fig.~\ref{fig:e_perp}, the parallel component of the effective permittivity $\varepsilon_\parallel$ is positive (at spatial frequencies for which the effective-medium model is valid). Therefore, the transition from the ellipsoid to solid hyperboloid in the case $a/b=0.05$ occurs at $k_0b\approx 25.4$, and in the case $a/b=0.08$ -- at $k_0b\approx 25.2$. 
Unlike our previous subsection, we do not depict here the dynamics of the topological transition because it is similar to that in the stacked hyperbolic metamaterial (see Fig.~3 of \cite{chebykin}). 

At frequencies $k_0b\approx 26$ ($a/b=0.08$) and $k_0b=25.7$ ($a/b=0.05$), the solid hyperboloid is again substituted by an ellipsoid. This topological transition is also characterized by $\Re(\varepsilon_\perp)=0$ and implies the maximum of the Purcell factor. However, as we will see below, in practical lossy media these two maxima do not exist separately. The ratio $\varepsilon_\parallel/\varepsilon_\perp$ may be large enough at the local minimum of $\varepsilon_\perp(\omega)$ which occurs at a Lorentz's resonance of the effective medium without crossing the zero.
Then the Purcell factor also experiences a resonance though the dispersion keeps elliptic. This regime (mimicking the 2d topological transition by the minimum of $\varepsilon_\perp(\omega)$) will be also considered in the next section.

Since both topological transitions in a WM -- that of the first type and that of the second type -- correspond to the resonances of the Purcell factor, an interesting problem arises: to approach these two resonances and, overlapping them, in order to obtain a broadband and strong Purcell effect where the values of the Purcell factor would dramatically exceed those obtained in Refs.~\cite{krishnamoorthy1} and \cite{chebykin} for stacked metamaterials and even the values obtained in \cite{poddubny3} for perfectly conducting WM.
In the following section we design the needed WM and theoretically reveal the expected effect for its  finite-size sample.

\section{Optimization of the wire medium sample}
\label{sec:results}

We start from the volume fraction of nanowires $f_{\rm{v}}=0.0804$. This value was used in our previous study \cite{new} in order to demonstrate the topological transition of the first type.  
In this case, Eq.~\ref{eq:condition} gives (assuming that the optical losses are sufficiently small) the relative dielectric constant of the wires $\Re({\varepsilon_{\rm{r}}})=-11.4$. This dielectric response is modestly negative. Polaritonic materials like lithium tantalate (LiTaO$_3$) can provide such response at infrared frequencies. For this range, the chosen volume fraction may correspond to a lattice constant $b=$200 nm and wire radius $a=$32 nm which are much smaller than the wavelength. Therefore, the effective-medium model is really accurate and reliable to use. In the present work, we again choose lithium tantalate (LiTaO$_3$) as the material of the wires like the previous study \cite{new}.

The dielectric function of LiTaO$_3$ is described by the Drude-Lorentz model \cite{kittel1}:
\begin{equation}
\varepsilon_{\rm{r}}=\varepsilon_\infty\left[1+\displaystyle\frac{\omega_{\rm{L}}^2-\omega_{\rm{T}}^2}{\omega_{\rm{T}}^2-\omega^2+j\omega\gamma}\right],
\end{equation}
in which $\omega_{\rm{T}}/2\pi=26.7$ THz, $\omega_{\rm{L}}/2\pi=46.9$ THz, and $\varepsilon_\infty=13.4$ \cite{schall1, crimmins1}. This frequency dependence is in a very good agreement with the experiment. Thirty nine (39) THz is the frequency where the real part of the relative dielectric constant is equal to $\Re(\varepsilon_{\rm{r}})=-11.4$, that is the target at the present stage.

Next, Eq.~\ref{eq:pz} determines the condition for the second type of topological transition. Since $f_{\rm{v}}\ll 1$ the pole and zero of the perpendicular permittivity are located closely on the frequency axis.
Therefore $P\approx Z\approx -1$ and, consequently, $\varepsilon_{\rm{r}}\approx -1$. This value corresponds to the frequency range 45.5--46.0 THz. Thirty nine (39) THz and 46 THz are rather distant frequencies, and the
corresponding resonances of the Purcell factor do not overlap. However, on the present stage it is worth to show that the radiation of the parallel dipole really experiences both these resonances. 

\begin{figure}[t!]\centering
\includegraphics[width=8cm]{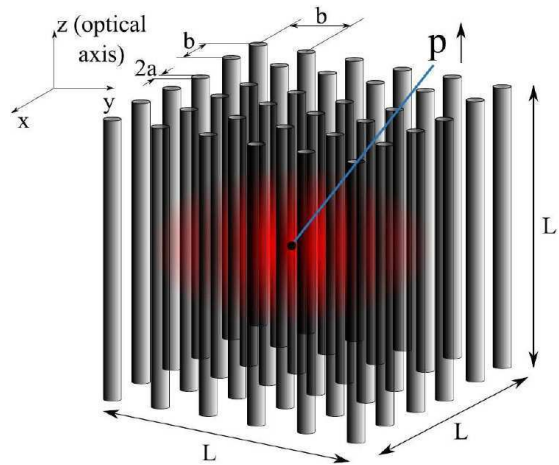}
\caption{Sub-wavelength parallel electric dipole in the cubic-shape sample of wire medium. The color spot illustrates the radiation which attenuates mainly due to optical losses. }
\label{fig:dipole}
\end{figure}

In the following, we report the numerical results for the Purcell factor of a finite sample of optimized WM for subwavelength electric dipoles either parallel or perpendicular. 
We performed full-wave simulations using the CST Microwave Studio simulator. For reliability we applied two solvers (FDTD and frequency domain ones) and attained the coincidence.
In our simulations, the sample of the WM is always of cubic shape with same internal geometry though its dimensions were varied. A subwavelength electric dipole (the same as in our work \cite{new}) 
is positioned symmetrically -- centered within the gap between four central nanowires. This location is nearly equivalent to the averaging over different positions of the dipole \cite{new}. 
The simulated structure is shown in Fig.~\ref{fig:dipole}. 

We calculated two types of the Purcell factor. The first one is the usual (or full) Purcell factor i.e. an overall enhancement of the radiated power of the given dipole moment in the environment different from free space. The second factor is the so-called radiative Purcell factor that is, indeed, the enhancement of the power radiated by the given dipole into free space. Definitely, for an optically large sample of lossy WM the second factor is much smaller because the most part of radiation is absorbed in the medium. We prove by extended simulations that the obtained values for the full Purcell factor 
can be identified with those of the Purcell factor of the infinite WM. As to the radiative Purcell factor, it is not so, and the results depend on the sample size. 

\subsection{Parallel dipole emitter}
\begin{figure}[htb!]\centering
\includegraphics[width=8cm]{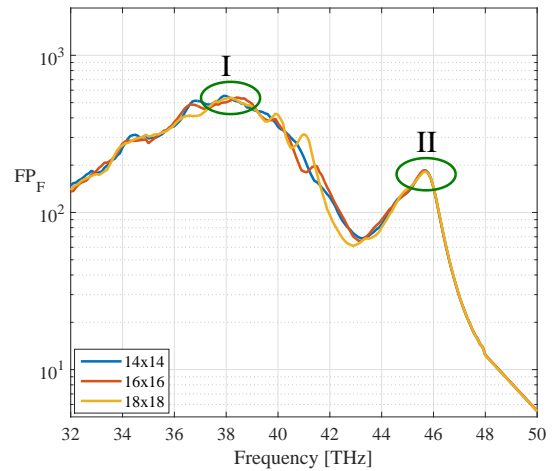}
\caption{Full Purcell factor versus frequency for different sizes of wire-medium sample. Here, the radius of each wire and the lattice constant of the medium are 32 nm and 200 nm, respectively. The emitter is parallel to the optical axis.}
\label{fig:FPF1}
\end{figure}

Geometry and orientation of the dipole is shown in Fig.~\ref{fig:dipole}. The full Purcell factor versus frequency is presented in Fig.~\ref{fig:FPF1}.
The maxima of the Purcell factor have nothing to do with the finite size of the sample and can be identified with the Purcell factor of an inifnite WM. We checked the absence of dimensional resonances simulating the sample with three different sizes corresponding to $14\times14$, $16\times16$ and $18\times18$ nanowires. Since we keep the cubic shape of the sample, the length of the wires was in these simulations equal to 2.664 $\mu$m, to 3.064 $\mu$m and to 3.464 $\mu$m, respectively. Two peaks of the full Purcell factor located at 38 THz and 46 THz (i.e. very close to values predicted by our theory) are kept for all three sizes of the sample. Also, the values of the Purcell factor at these maxima was kept. At the first topological resonance, the full Purcell factor is nearly equal to 500, and at the second one it is equal to 180.
This stability means that dimensional (Fabry-Perot) resonances at the frequencies of our interest are absent. Really, calculations of the optical decay using the effective-medium model shows that these resonances 
are damped by losses even for a structure with $12\times12$ nanowires. 

%The reason of the small discrepancy for the first topological transition (38 against 39 THz) is the effect of losses ignored in our estimations. Removing losses in our simulations one may approach this resonance to 39 THz. %However, parasitic dimensional resonances arise in the lossless case and the interpretation of the simulation data becomes ambiguous.

\begin{figure}[t!]\centering
\subfigure[]{
\includegraphics[width=8cm]{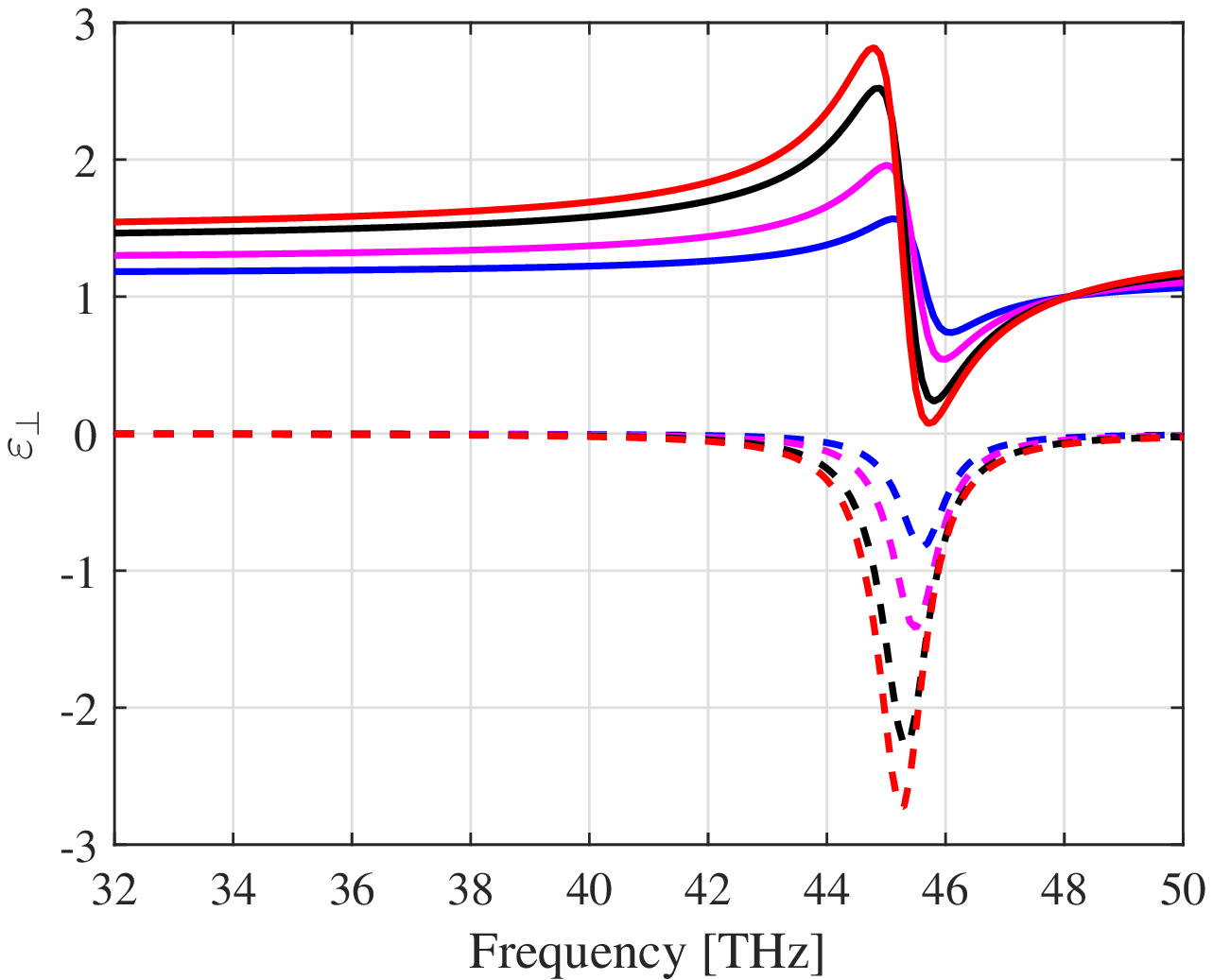}
\label{fig:pertransverse}}
\subfigure[]{
\includegraphics[width=8cm]{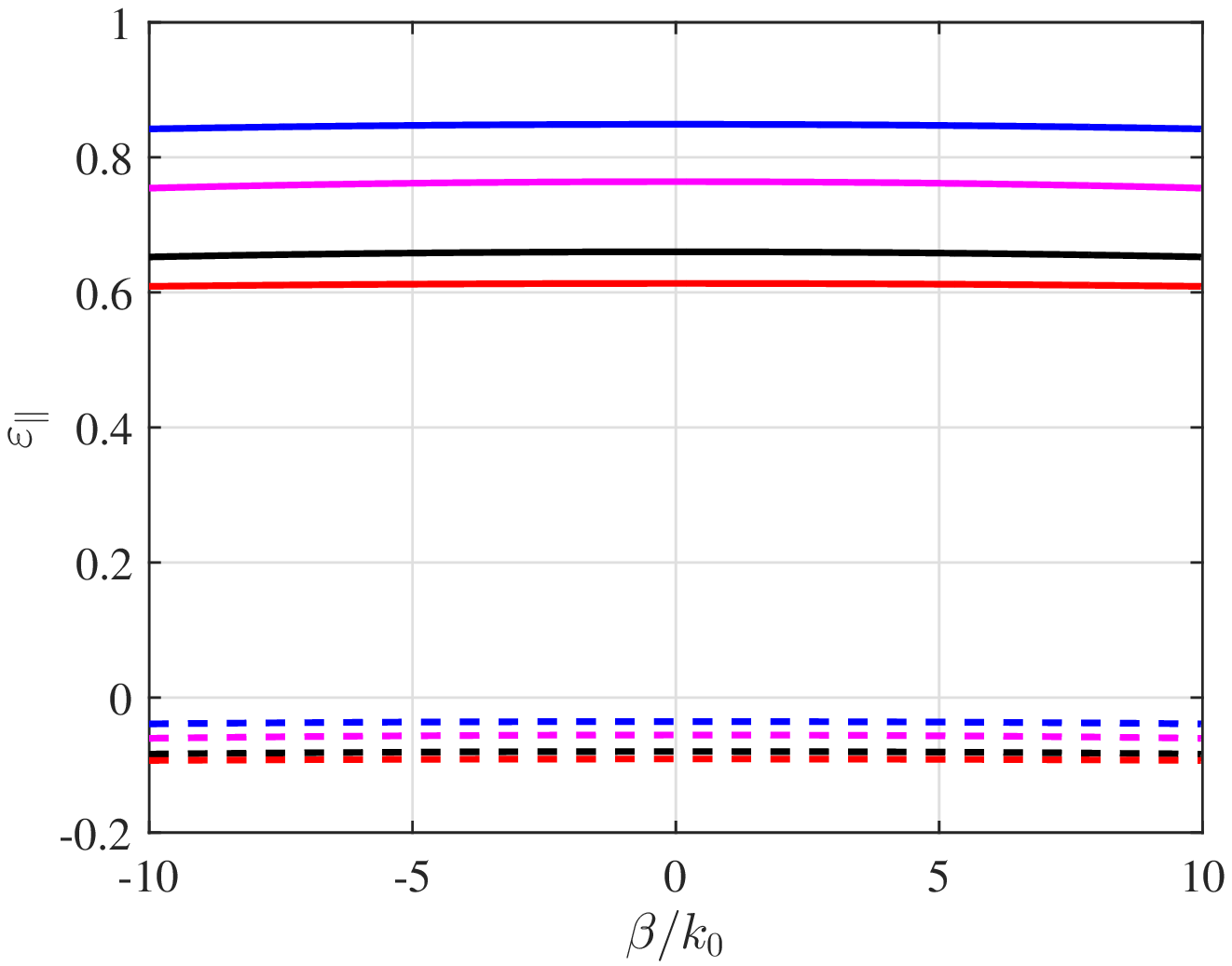}
\label{fig:perlongitudinal}}
\caption{a)--Perpendicular component of permittivity with respect to the frequency for lithium tantalate wire medium. b)--Parallel component of permittivity with respect to the normalized wave vector at $f=45.9$ THz. The solid lines show the real part and the dashed lines determine the imaginary part. The blue, magenta, black and red colors correspond to the ${\rm{radius}}=a=32$ nm, ${\rm{radius}}=1.25a$, ${\rm{radius}}=1.5a$ and ${\rm{radius}}=1.6a$, respectively.}
\end{figure}

As to the second topological transition, it is important to analyze the resonance of the perpendicular permittivity, which is different from that analyzed in the previous subsection due to much higher optical losses in lithium tantalate. The frequency dispersion of $\varepsilon_\perp$ is shown in Fig.~\ref{fig:pertransverse}. The blue curve corresponds to the initial value $a=$32 nm and to Fig.~\ref{fig:FPF1}. Due to high losses, the real part of $\varepsilon_\perp$ does not change the sign at the resonance. At 46 THz $\Re(\varepsilon_\perp)$ is minimal and smaller than unity. This is sufficient to mimic the regime $\varepsilon_\perp=0$ which offers the second topological transition. As shown in Fig.~\ref{fig:pertransverse}, increasing the radius of the wires (the lattice constant is fixed) causes a stronger resonance which makes $\Re(\varepsilon_\perp)$ become closer to zero. If $\Re(\varepsilon_\perp)$ crosses the zero axis, then the zeros of $\Re(\varepsilon_\perp)$ occur at frequencies which are close to one another and in this range the WM is not very lossy. Really, in Fig.~\ref{fig:pertransverse}, one may see that the maximum of the lossy factor $|\Im(\varepsilon_\perp)|$ occurs at a frequency slightly lower than the band in which $\Re(\varepsilon_\perp)$ is minimal. Therefore the increase of $a$ is favorable for the second resonance of the Purcell factor. Indeed, increasing the fraction of nanowires we risk to lose the first topological transition. Therefore, for given $b$ there must be optimal values for the wire radius. This optimum is to be found.  

\begin{figure}[t!]\centering
\subfigure[]{
\includegraphics[width=8cm]{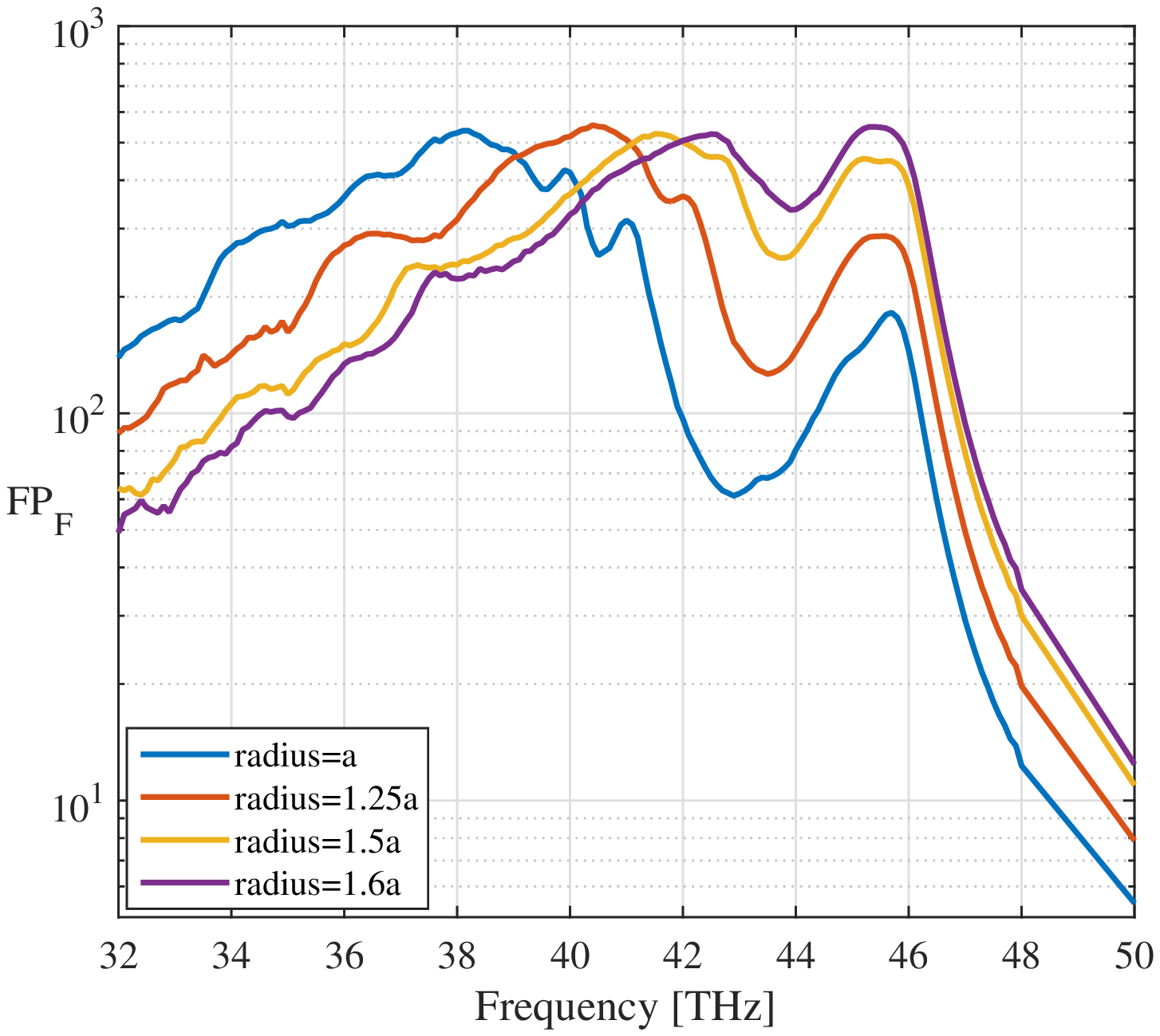}
\label{fig:FPF2}}
\subfigure[]{
\includegraphics[width=8cm]{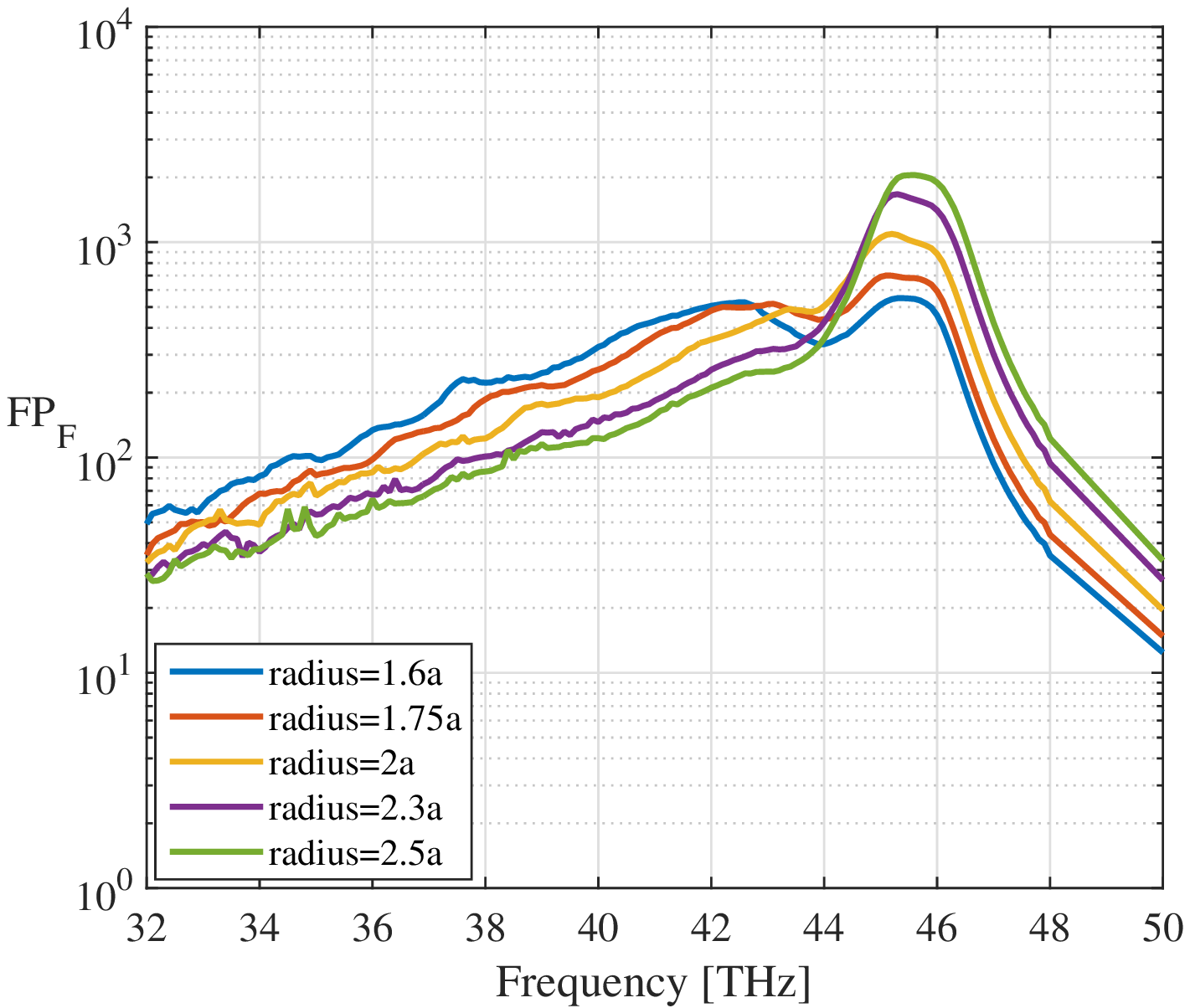}
\label{fig:FPF3}}
\caption{Full Purcell factor ($FP_{\rm{F}}$) with respect to the frequency for several values of the wires' radii. The lattice constant is $200$ nm, and the initial radius is $a=32$ nm. The emitter is parallel to the optical axis and the sample includes $18\times18$ nanowires.}
\end{figure}

In accordance to the insight that the Purcell factor of a very anisotropic medium is determined by the ratio $\varepsilon_\parallel$/$\varepsilon_\perp$, we inspected also the parallel component of the effective permittivity. Figure.~\ref{fig:perlongitudinal} shows $\varepsilon_\parallel$ versus normalized parallel wave number at 45.9 THz. We see that at this frequency, $\Re(\varepsilon_\parallel)$ is approximately uniform versus $\beta$, and it is a positive value in between 0.6 and 0.85 for wire radius varying within 60\% range. The increase of the wire radius is important for the perpendicular permittivity where it allows the 2nd topological transition when the wire radius becomes larger than $1.6a$. For the parallel permittivity in this frequency range the wire radius is not very important. 

The optimization of the WM implies that we need to approach two resonances of the Purcell factor and to enhance the second resonance compared to the values achieved in Fig.~\ref{fig:FPF1}. Increasing the radius of nanowires we pass through its optimal value. In Fig.~\ref{fig:FPF2}, we show the simulated Purcell factor for wire radii increasing from $a=32$ nm to $1.6a=51.2$  nm. With these values of the wire radius we only mimic the second topological transition. However, we keep the 1st one. The Purcell factor corresponding to this first resonance in Fig.~\ref{fig:FPF2} keeps nearly the same for different wire radii. However, the resonance frequency moves versus the wire radius and approaches to 46 THz. Two resonances start to overlap when the radius exceeds $a$. The best case corresponds to the wire radius $1.6a=51.2$ nm when $FP_{\rm{F}}>300$ at 40--46 THz. The two resonances can be still distinguished but are almost identical and the dip between them is small. If the target is the broadest frequency band in which Purcell factor exceeds 300, the best result corresponds in this plot to $1.6a$.

\begin{figure}[t!]\centering
\subfigure[]{
\includegraphics[width=8cm]{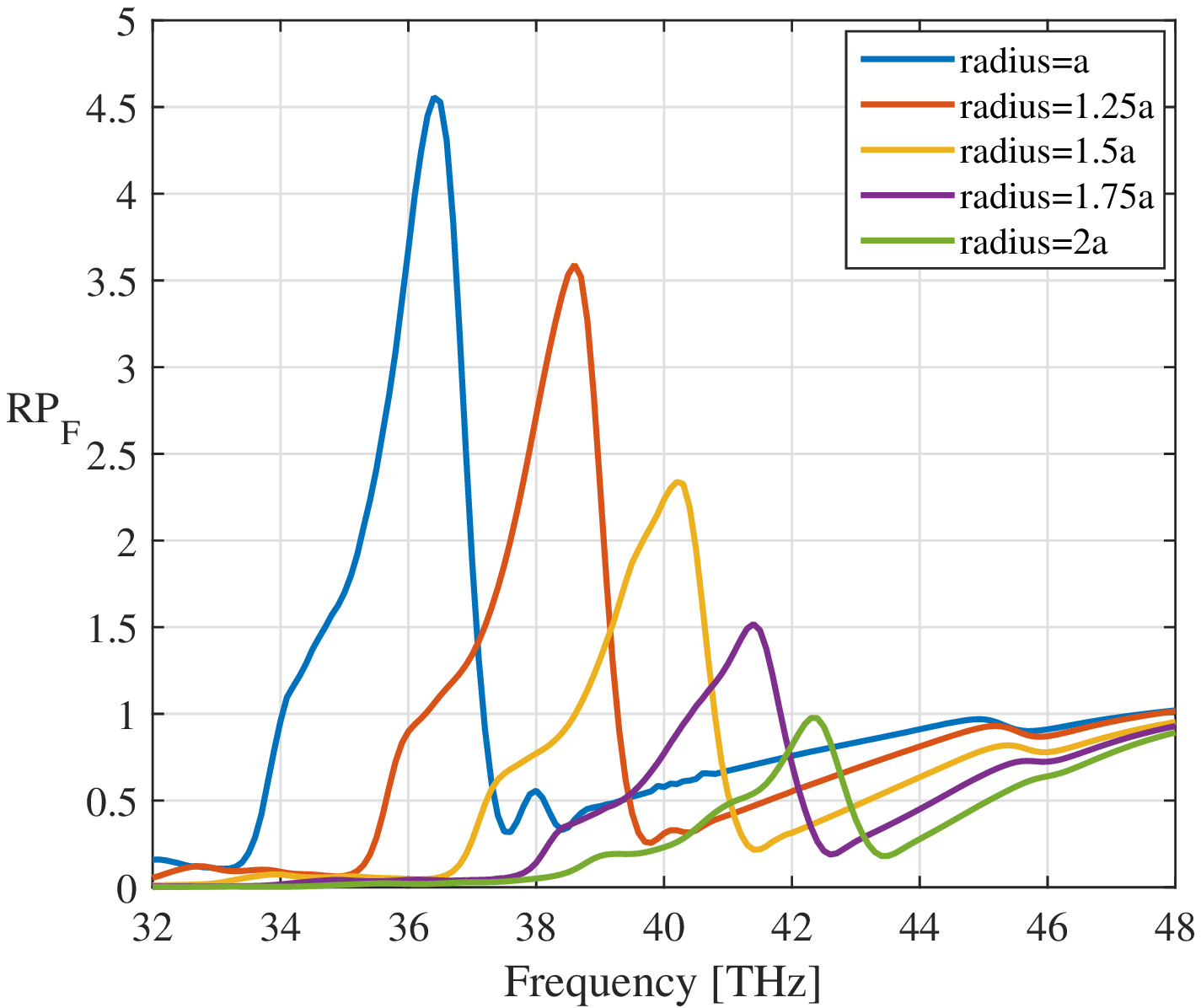}
\label{fig:RPF1}}
\subfigure[]{
\includegraphics[width=8cm]{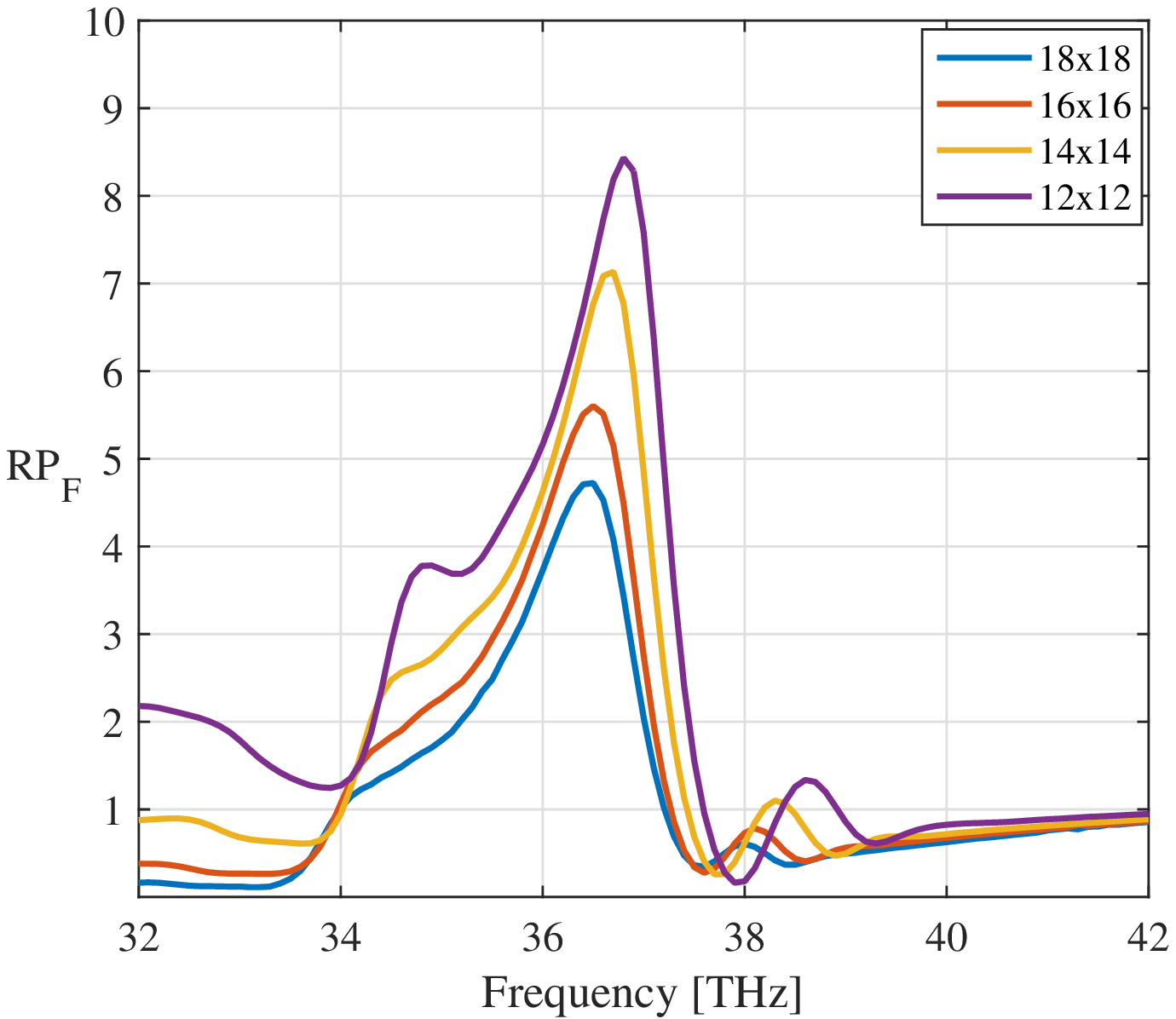}
\label{fig:RPF2}}
\caption{Radiative Purcell factor ($RP_{\rm{F}}$) with respect to the frequency. (a)--$RP_{\rm{F}}$ for several values of the wires' radii. The lattice constant is $200$ nm, and the initial radius is $a=32$ nm. The emitter is parallel to the optical axis and the sample includes $18\times18$ nanowires. (b)--$RP_{\rm{F}}$ for different sizes of the sample. Here, the lattice constant is as before and the wire radius is 32 nm.}
\end{figure}

However, the wire radius 51.2 nm is also the lower bound of radii for which $\Re(\varepsilon_\perp)$ changes the sign, and allows the second-type topological transition to really occur. Therefore, the Purcell factor at this resonance grows versus wire radii as it is shown in Fig.~\ref{fig:FPF3}. As seen in this figure, the first resonance becomes sacrificed if the wire radius exceeds $1.75a$. On the other hand, Purcell factor at the second resonance still increases versus the radius and the best result corresponds to the radius $2.5a=$80 nm. In this case the maximal Purcell factor is equal $2\cdot 10^3$ and this value is by an order of magnitude higher than the values obtained for WM in \cite{new} and for stacked media in \cite{chebykin}. This amazing radiation enhancement is revealed for a parallel dipole which was previously considered as a very inefficient emitter for WM.

It is worth to notice that at the second resonance the ordinary (TE) mode may contribute into the Purcell factor. For small fraction volumes and at lower frequencies the perpendicular component of the effective permittivity is approximately equal to unity implying $P_{\rm{F}}\,({\rm{TE}})\approx 1$. However, at the frequency of the second resonance $\Re(\varepsilon_\perp)\approx 0$ or even $\Re(\varepsilon_\perp)= 0$. Therefore, at this frequency the TE mode may also play a noticeable role in the Purcell effect. In this work, we, however, do not share out the contribution of the TE-modes into Purcell factor and leave this question for our next paper.

Our next plot -- Fig.~\ref{fig:RPF1} -- shows the radiative Purcell factor. It is still plotted versus frequency for different wire radii. Comparing Figs.~\ref{fig:FPF2} and \ref{fig:FPF3} to Fig.~\ref{fig:RPF1}, we see that 
the radiative Purcell factor feels the first topological transition (although keeps very small compared to the full one).
However,  the frequency where $RP_{\rm{F}}$ is maximal is determined in a contest of the resonance for the full Purcell factor (total radiated power) and the requirement of the impedance matching at the WM interfaces. 
This is why the frequency of this maximum varies when the wire radius increases (and, respectively, the wave impedance of the WM  changes).

Notice that around the frequency of the second topological transition the radiative Purcell factor is smaller than unity due to the strong impedance mismatch (total internal reflection). Though the total radiated power is maximal in this frequency band the radiation  into free space is lower than that of a dipole in absence of the WM. This is the result of the almost total confinement of radiation in the sample whose effective permittivity is smaller than unity (see Figs.~\ref{fig:pertransverse} and ~\ref{fig:perlongitudinal}). 

Figure.~\ref{fig:RPF2} clearly shows the absence of dimensional resonances around the
frequency of the first transition representing radiative Purcell factors for different sizes of the sample. Approximately all the peaks happen at the same frequency and peak value decreases versus the sample size in accordance to growing attenuation across the sample. In general, all our plots show that the full Purcell factor simulated for a finite sample can be really identified with that of the infinite WM. On the contrary, the radiative Purcell factor is determined by the losses in the sample and by the mismatch at the interface. At the first resonance the matching is quite good and the size reduction increases the resonant value of the
radiative Purcell factor. However, the size of the sample can not be very small in order to keep the properties of WM. Practically, it implies that the optimal sample size corresponds to $12\times12$ nanowires.
Still, in this best case 99\% of radiated power are dissipated in the WM, and the radiative Purcell factor attains only 8.3. 

\subsection{Perpendicular dipole}
\begin{figure}[htb!]\centering
\subfigure[]{\includegraphics[width=8cm]{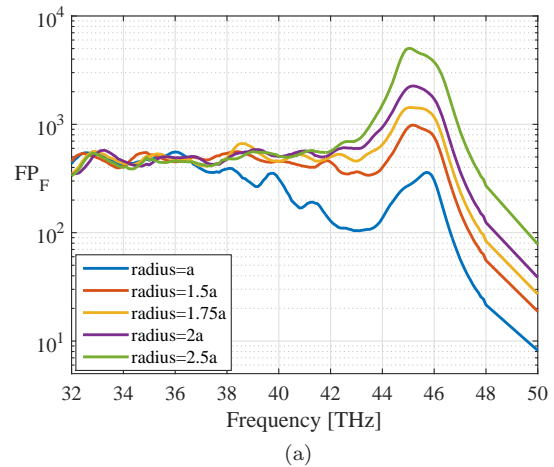}
\label{fig:FPF4}}
\subfigure[]{\includegraphics[width=8cm]{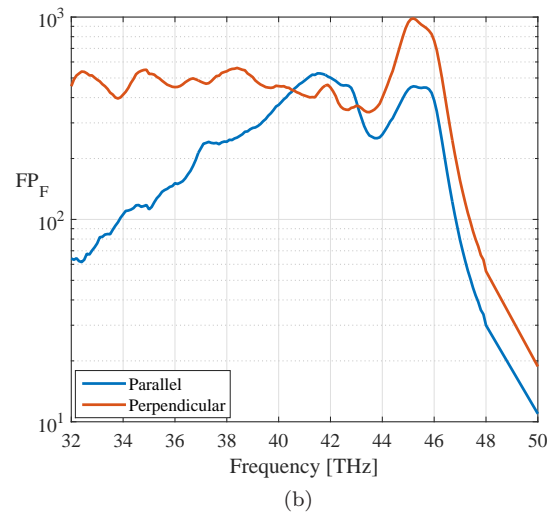}
\label{fig:FPF5}}
\caption{(a)--Full Purcell factor for several values of the wires' radii. The emitter is perpendicular to the optical axis. Here, $a=32$ nm and $b=200$ nm. The sample includes $18\times18$ nanowires. (b)--Full Purcell factor for two different orientations of the emitter in the case of ${\rm{radius}}=1.5a$.}
\end{figure}

Here, we report the results obtained for a dipole perpendicular to the optical axis and compare these results with those for a parallel dipole. Figure~\ref{fig:FPF4} shows the full Purcell factor as a function of frequency for different radii of the wires. The first topological transition does not have impact to the radiation of the perpendicular dipole. This result is in line with our speculations in \cite{new}. The Purcell factor remains approximately uniform around the frequency of the first transition and its value in the order of magnitude matches the predictions of \cite{poddubny3}.
Between 45 THz and 46 THz the Purcell factor has a local maximum which corresponds to the second type of the transition. As seen, the Purcell factor at resonant frequency increases as the wire radius increases.

To see clearer the difference between the parallel and perpendicular dipoles, we have shown the full Purcell factor for these two dipole orientations in Fig.~\ref{fig:FPF5} when the wire radius is fixed as $1.5a=$48 nm. At the first transition the Purcell factor for the parallel dipole is higher. At the second transition the perpendicular dipole radiates more. For values of the wire radius within $a$ and $2a$ there is the band in which the parallel and perpendicular dipole radiate approximately the same power. For the radius $1.5a$ it is the band 40--44 THz. Emitters will all orientations of their polarization radiate in this band nearly the same power. This is the manifestation of the nearly isotropic Purcell factor which is a surprising effect for a very anisotropic medium. This isotropy and the wide band of the Purcell factor allows us to suggest another application of the Purcell effect, namely the radiative cooling of microscopic objects. Thermal radiation is usually isotropic and broadband. Very high values of the broadband and isotropic Purcell factor imply the very efficient radiation of the excessive heat into WM, that may be important for micro- and nanolasers and other active devices which can be incorporated into a sample of polaritonic WM.

\section{Conclusions}
\label{sec:conclusions}
We have theoretically studied two different topological transitions of the dispersion that may happen in the same wire medium at different frequencies. Previously, we already revealed one of these transitions for a special case when the WM is performed of polaritonic nanowires. In this work we have found the second type of topological transition which appears at higher frequencies where the perpendicular to the optical axis
component of the effective permittivity tensor crosses zero. This regime can be also mimicked by the resonance of the perpendicular permittivity when its real part attains a positive value much smaller than unity.
Both these topological transitions correspond to resonances of the Purcell factor for a parallel dipole. We have shown that optimizing the WM, one may bring the two resonances to one another. Overlapping the corresponding resonance bands we achieve the huge and broadband enhancement of the power radiated by a parallel dipole. We confirmed our theoretical expectations by extensive simulations. The two types of resonances for the Purcell factor corresponding to two topological transitions were specified for a sub-wavelength dipole located at the center of a finite-size sample of a polaritonic WM. We have studied the radiative Purcell factor and explained its significant difference from the full one. Finally, we compared and discussed the frequency behavior of the Purcell factor for dipoles oriented in parallel to the optical axis and perpendicularly. We have found the regime of the resonant isotropic Purcell factor. The physical effects theoretically revealed in this work allows the broadband and strong radiation enhancement which can be applied in optical nanosensing and in radiative cooling of small objects.

We claimed all reported effects for WM performed of lithium tantalate. However, our effects can be also engineered with other polaritonic materials, e.g. crystalline or polycrystal silicon carbide (SiC). The work \cite{mahi1} reported the fabrication of some polaritonic WM for the infrared operation grown in a dielectric matrix. Perhaps, there are no limitations for fabrication of polaritonic WM with free-standing parts (as in our simulations). However, we believe that main results of the present work can be reproduced for nanowires in a dielectric matrix with host permittivity larger than unity. Free space as a host medium is chosen in this work only for simplicity of calculations.

\section{Acknowledgments}
This work was supported by Aalto ELEC Doctoral School. The authors of the paper would like to thank S.~A.~Tretyakov for an interesting and useful discussion.    

%%%%%%%%%%%%%%%%%%%%%%%%%%%%%%%%%%%%%%%%%%%%%%%%%%%%%%%%%%%%%%%%%%%%%%%%%%%%%%%%%%%


\begin{thebibliography}{00}

\bibitem{simovski2}
C.~R.~Simovski, P.~A.~Belov, A.~V.~Atrashchenko and Y.~S.~Kivshar,
Wire metamaterials: physics and applications,
Adv.~Mater.~\textbf{24}, 4229--4248 (2012).

\bibitem{belov1}
P.~A.~Belov, Y.~Zhao, S.~Tse, P.~Ikonen, M.~G.~Silveirinha, C.~R.~Simovski, S.~Tretyakov, Y.~Hao and C.~Parini,
Transmission of images with subwavelength resolution to distances of several wavelengths in the microwave range,
Phys.~Rev.~B~\textbf{77}, 193108 (2008).

\bibitem{casse}
B.~D.~F.~Casse, W.~T.~Lu, Y.~J.~Huang, E.~Gultepe, L.~Menon, S.~Sridhar,
Super-resolution imaging using a three-dimensional metamaterials nanolens,
Appl.~Phys.~Lett.~\textbf{96}, 023114 (2010).

\bibitem{simovski1}
I.~S.~Nefedov and C.~R.~Simovski,
Giant radiation heat transfer through micron gaps,
Phys.~Rev.~B~\textbf{84}, 195459 (2011).

\bibitem{mirmoosa1}
M.~S.~Mirmoosa and C.~R.~Simovski,
Micron-gap thermophotovoltaic systems enhanced by nanowires,
Photon.~Nanostruct.~Fundam.~Appl.~\textbf{13}, 20--30 (2015).

\bibitem{burghignoli1}
P.~Burghignoli, G.~Lovat, F.~Capolino, D.~R.~Jackson and D.~R.~Wilton,
Directive leaky-wave radiation from a dipole source in a wire-medium slab,
IEEE~Trans.~Antennas Propag.~\textbf{56}, 1329--1339 (2008).

\bibitem{poddubny3}
A.~N.~Poddubny, P.~A.~Belov and Y.~S.~Kivshar,
Purcell effect in wire metamaterials,
Phys.~Rev.~B.~\textbf{87}, 035136 (2013).

\bibitem{smith1}
D.~R.~Smith and D.~Schurig,
Electromagnetic wave propagation in media with indefinite permittivity and permeability tensors,
Phys.~Rev.~Lett.~\textbf{90}, 077405 (2003).

\bibitem{poddubny1}
A.~Poddubny, I.~ Irosh, P.~Belov and Y.~Kivshar,
Hyperbolic metamaterials,
Nature~Photonics~\textbf{7}, 948--957 (2013).

\bibitem{poddubny2}
A.~N.~Poddubny, P.~A.~Belov and Y.~S.~Kivshar,
Spontaneous radiation of a finite-size dipole emitter in hyperbolic media,
Phys.~Rev.~A~\textbf{84}, 023807 (2011).

\bibitem{kidwai1}
O.~Kidwai, S.~V.~Zhukovsky and J.~E.~Sipe,
Dipole radiation near hyperbolic metamaterials: applicability of effective-medium approximation,
Opt.~Lett.~\textbf{36}, 2530--2532 (2011).

\bibitem{purcell1}
E.~M.~Purcell,
Spontaneous emission probabilities at radio frequencies,
Phys.~Rev.~\textbf{69}, 681 (1946).

\bibitem{novotny1}
L.~Novotny and B.~Hecht, \emph{Principles of Nano-Optics} (Cambridge University Press, Cambridge, UK, 2006).

\bibitem{sauvan1}
C.~Sauvan, J.~P.~Hugonin, I.~S.~Maksymov and P.~Lalanne,
Theory of the spontaneous optical emission of nanosize photonic and plasmon resonators,
Phys.~Rev.~Lett.~\textbf{110}, 237401 (2013).

\bibitem{pelton1}
M.~Pelton,
Modified spontaneous emission in nanophotonic structures,
Nature~Photonics~\textbf{9}, 427--435 (2015).

\bibitem{tam1}
F.~Tam, G.~P.~Goodrich, B.~R.~Johnson and N.~J.~Halas,
Plasmonic enhancement of molecular fluorescence,
Nano~Lett.~\textbf{7}, 496--501 (2007).

\bibitem{anger1}
P.~Anger, P.~Bharadwaj and L.~Novotny,
Enhancement and quenching of single-molecule fluorescence,
Phys.~Rev.~Lett.~\textbf{96}, 113002 (2006).

\bibitem{jacob1}
Z.~Jacob, I.~I.~Smolyaninov and E.~E.~Narimanov,
Broadband Purcell effect: radiative decay engineering with metamaterials,
Appl.~Phys.~Lett.~\textbf{100}, 181105 (2012).

\bibitem{Krasnok}
A.~E.~Krasnok, A.~P.~Slobozhanyuk, C.~R.~Simovski, S.~A.~Tretyakov, A.~N.~Poddubny,
A~.E.~Miroshnichenko, Y.~S.~Kivshar and P.~A.~Belov,
An antenna model for the Purcell effect,
Sci.~Rep.~\textbf{5}, 12956 (2015).


\bibitem{krishnamoorthy1}
H.~N.~S.~Krishnamoorthy, Z.~Jacob, E.~Narimanov, I.~Kretzschmar and V.~M.~Menon,
Metamaterial based broadband engineering of quantum dot spontaneous emission,
e-print arXiv:0912.2454v1 [physics.optics].



\bibitem{krishnamoorthy2}
H.~N.~S.~Krishnamoorthy, Z.~Jacob, E.~Narimanov, I.~ Kretzschmar and V.~M.~Menon,
Topological transitions in metamaterials,
Science~\textbf{336}, 205--209 (2012).

\bibitem{yang}
X.~Yang, C.~Hu, H.~Deng, D.~Rosenmann, D.~A.~Czaplewski and J.~Gao,
Experimental demonstration of near-infrared epsilon-near-zero multilayer metamaterial slabs,
Opt.~Express~\textbf{21}, 23631--23639 (2013).


\bibitem{new}
M.~S.~Mirmoosa, S.~Yu.~Kosulnikov and C.~R.~Simovski, 
Unbounded spatial spectrum of propagating waves in a polaritonic wire medium,
Phys.~Rev.~B~\textbf{92}, 075139 (2015).


\bibitem{silveirinha1}
M.~G.~Silveirinha,
Nonlocal homogenization model for a periodic array of $\epsilon$-negative rods,
Phys.~Rev.~E~\textbf{73}, 046612 (2006).

\bibitem{marcuvitz1}
L.~B.~Felsen and N. Marcuvitz, \emph{Radiation and Scattering of Waves} (IEEE Press, New York, 1994).


\bibitem{chebykin}
A. V. Chebykin, A. A. Orlov, A. S. Shalin, A. N. Poddubny, and P. A. Belov,
Strong Purcell effect in anisotropic epsilon-near-zero metamaterials,
Phys. Rev. B \textbf{91}, 205126 (2015).


\bibitem{kittel1}
C.~Kittel, \emph{Introduction to Solid State Physics} (Wiley, New York, 1976).

\bibitem{schall1}
M.~Schall, H.~Helm and S.~R.~Keiding,
Far infrared properties of electro-optic crystals measured by THz time-domain spectroscopy,
Int.~J.~Infrared Millim.~Waves~\textbf{20}, 595--604 (1999).

\bibitem{crimmins1}
T.~F.~Crimmins, N.~S.~Stoyanov and K.~A.~Nelson,
Heterodyned impulsive stimulated Raman scattering of phonon–polaritons in LiTaO3 and LiNbO3,
J.~Chem.~Phys.~\textbf{117}, 2882--2896 (2002).


\bibitem{mahi1}
Mahi~R.~Singh,
Polaritonics in nanowires made from dispersive materials,
Phys.~Rev.~B~\textbf{80}, 195303 (2009).


\end{thebibliography}
\end{document}